\documentclass[epj]{svjour}
\usepackage{latexsym}
\usepackage{graphics}
\usepackage{amssymb}
\usepackage{amsmath}
\usepackage{widetext}
\usepackage{tablefootnote}
\usepackage{xcolor}

\begin{document}
\title{Cross-section of the ${^{\rm nat}\rm{Ni}}(\gamma,{\rm p} x\rm n)^{58}\rm{Co}$ reaction \\
	at the bremsstrahlung end-point energy of 35--94~MeV}
\author{I.S. Timchenko\inst{1,}\inst{2,}\thanks{\emph{Corresponding author:}\\
		iryna.timchenko@savba.sk;\\  timchenko@kipt.kharkov.ua}%
\and O.S. Deiev\inst{2}
\and S.M. Olejnik\inst{2}
\and S.M. Potin\inst{2}
\and V.A. Kushnir\inst{2}
\and V.V. Mytrochenko\inst{2} 
\and S.A.~Perezhogin\inst{2} 
\and A. Herz\'{a}\v{n}\inst{1}  
}                     
%
%
\institute{Institute of Physics, Slovak Academy of Sciences, SK-84511 Bratislava, Slovakia
\and NSC "Kharkov Institute of Physics and Technology", National Academy of Sciences of Ukraine, 1 Academichna Str., Kharkiv, 61108, Ukraine
}%
\date{Received: date / Revised version: date}
%
\abstract{
	The photoproduction of the $^{58}\rm{Co}$ nuclei on ${^{\rm nat}\rm{Ni}}$ was studied using the induced $\gamma$-activity method and off-line  $\gamma$-ray spectrometric technique. The experiment was performed at the electron linear accelerator LUE-40 NSC KIPT, Ukraine. The total flux-averaged cross-section $\langle{\sigma(E_{\rm{\gamma max}})}\rangle$ for the ${^{\rm nat}\rm{Ni}}(\gamma,{\rm p} x\rm n)^{58}\rm{Co}$ reaction has been measured in the range of bremsstrahlung end-point energies $E_{\rm{\gamma max}}$ = 35--94~MeV. The obtained $\langle{\sigma(E_{\rm{\gamma max}})}\rangle$ were compared with theoretical estimates. The theoretical values of $\langle{\sigma(E_{\rm{\gamma max}})}\rangle_{\rm th}$ were calculated using the partial cross-sections $\sigma(E)$ from the TALYS1.96 code for different level density models and gamma strength functions. 
} 
\authorrunning{I.S. Timchenko, O.S. Deiev, S.N. Olejnik, S.M. Potin, ...}
\titlerunning{Cross-section of the ${^{\rm nat}\rm{Ni}}(\gamma,{\rm p} x\rm n)^{58}\rm{Co}$ reaction at ...}
\maketitle

\section{Introduction}
\label{intro}
Data on photonuclear reactions are necessary for studying fundamental problems in various disciplines, e.g. astrophysics, theoretical description of the structure of the atomic nucleus, the mechanism of nuclear reactions, competition between statistical and direct reaction channels, etc. In addition, these data hold significant value for both medical and applied physics research, contributing to advancements in pharmaceutical development \cite{a1}, the optimization of fast reactor designs \cite{a2}, and the refinement of accelerator-driven subcritical systems \cite{a3,a4}.

Despite the many efforts made to study photonuclear reactions, there is currently a lack of experimental data on multiparticle reactions and reactions with the emission of charged particles. It should also be noted that the available experimental data, obtained in the Saclay Nuclear Research Center (France) and Lawrence Livermore National Laboratory (United States), do not always agree in the energy region of Giant Dipole Resonance (GDR). For example, in the Atlas database \cite{1} it can be seen that the data for ($\gamma$,1n) reactions, differ systematically by 17\% in the maximum cross-sections.  
The systematic disagreements between cross sections of partial photoneutron ($\gamma$,1n) and ($\gamma$,2n) reactions for nuclei from $^{89}$Y up to $^{208}$Pb,  are discussed in Ref. \cite{Wolynec}.
For these reasons, previously measured data on photonuclear reaction cross-sections are re-analyzed to check for possible systematic errors, and new experimental studies of photoneutron reactions are  underway.
For example, cross-sections of reactions emitting up to 4 neutrons at energies up to 40~MeV have been measured at the NEWSubaru facility  \cite{Kavano}. In Refs. \cite{a11,a12,a13,a14} the cross-sections of photoneutron reactions weighted by the bremsstrahlung flux have been measured at energies above the GDR and up to the pion production threshold.

Note that in the case of multiparticle photoneutron reactions or reactions with charged particles/clusters in the output channel, the different variants of theoretical cross-sections have a more noticeable spread than the calculations for ($\gamma$,n) reaction. This leads to the need to use such data for testing the theoretical models. As an example, in the studies of photonuclear reactions on Cu \cite{Ku}, Ta \cite{Ta1,Ta2} or Al \cite{Al22Na}, calculations in the TALYS code \cite{8,9} for different level density models and their comparison with experimental cross-sections are shown.


In the literature, several studies reporting on the results of photonuclear reactions on natural nickel can be found. In Refs. \cite{25,26,27,28,30,31,32,33} data obtained using mono-energetic gamma-beam at energies of 12 -- 40 MeV are presented and discussed. Results of studies using bremsstrahlung beam with energy up to 100 MeV are reported in \cite{Ni-Ni,Ni-Co,34,35,Masum}.

The reaction aiming at $^{58}$Co production on $^{\rm nat}$Ni was
studied in \cite{34,35}. It was shown that the obtained experimental data differ significantly from the theoretical estimates calculated using the cross-sections from the TALYS code with default parameters.	

Recently, we performed a detailed study of photonuclear reactions on $^{\rm nat}$Ni using a bremsstrahlung beam with energies between 35--94 MeV. Resulting flux-averaged reaction cross-sections for production of $^{56,57}$Ni and cumulative reaction cross-sections for production of $^{55-57}$Co are reported in \cite{Ni-Ni,Ni-Co}.

Study presented here is a continuation of our experimental program aiming at investigation of photonuclear reactions on stable isotopes of nickel. Here, we report on the experimental results for the total bremsstrahlung flux-averaged cross-sections $\langle{\sigma(E_{\rm{\gamma max}})}\rangle$ for the $^{58}$Co production. The obtained data are compared both with the theoretical calculation performed using partial cross-sections $\sigma(E)$ from the TALYS1.96 code and with the experimental data available in the literature \cite{34,35}. 

\section{Experimental setup and procedure}
\label{sec:2}

The study of the  ${^{\rm nat}\rm{Ni}}(\gamma,{\rm p} x\rm n)^{58}\rm{Co}$ reaction
was performed using the $\gamma$-activation and off-line $\gamma$-ray spectrometry. A description of this technique can be found, e.g., in \cite{a11,Ku,44,45}. Below is a brief description of the setup used in the work and details of the experiment.

\begin{figure}[]
	\resizebox{0.51\textwidth}{!}{%
		\includegraphics{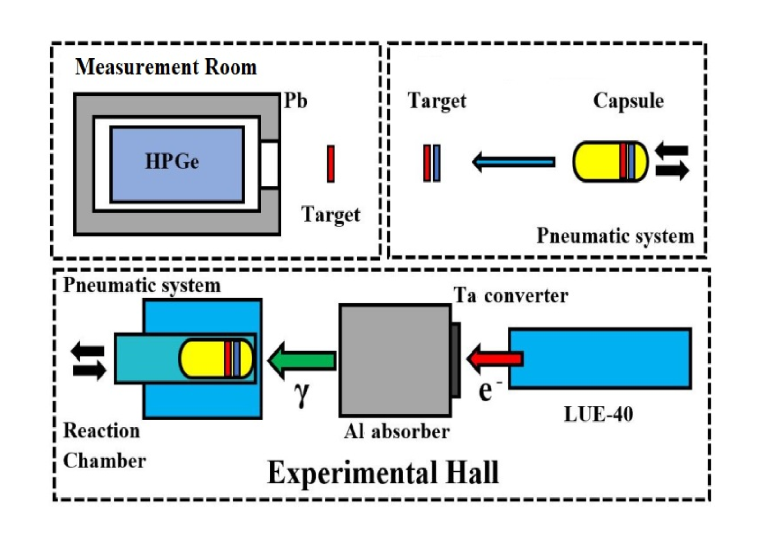}}
	\caption{Schematic block diagram of the experiment. The upper part shows the measurement room, where the irradiated target (red color) and the target-monitor (blue color) are extracted from the capsule and are arranged by turn before the HPGe detector for induced $\gamma$-activity measurements. The lower part shows the electron linear accelerator LUE-40, the Ta converter, Al absorber, explosure reaction chamber.}
	\label{fig1}
\end{figure}

The schematic block diagram of the experiment is presented in Fig.~\ref{fig1}. The experimental hall is shown in the lower part of the figure. A plate of natural tantalum, an aluminium cylinder, and a capsule with target and target-monitor are positioned on the axis of the electron beam of the linear accelerator LUE-40 of the Research and Development Center "Accelerator" of the National Science Centre "Kharkov Institute of Physics and Technologies", Ukraine \cite{46,47}. The tantalum plate was used as the target converter and was manufactured from a 1.05~mm thick natural tantalum plate, measuring 20 by 20~mm. To remove electrons from the bremsstrahlung $\gamma$-flux, the aluminum cylinder 100 mm in diameter and 150 mm in length, was used. For transporting the targets, placed in the aluminum capsule, to the place of irradiation and back for induced activity registration, the pneumatic tube transfer system was used. 

The upper part of the figure shows the measurement room, in which the detector is located, covered with a lead shield to reduce the natural background. The irradiated targets are delivered in this room, then the targets are extracted from the aluminum capsule and are transferred to the detector for the measurements. 

The induced  $\gamma$-activity of the irradiated targets was registered by the semiconductor HPGe detector Canberra GC--2018 with the resolutions of 0.8 and 1.8~keV (FWHM) for the energies 
$E_{\gamma}$ = 122 and 1332~keV, respectively. Its efficiency at 1332~keV was 20\% relative to the 
NaI(Tl) scintillator, 3 inches in diameter and 3 inches in thickness. The absolute registration efficiency of the HPGe detector was calibrated with a standard set of $\gamma$-ray radiation sources: $^{22}$Na, $^{60}$Co, $^{133}$Ba, $^{137}$Cs, $^{152}$Eu, $^{241}$Am. 

The $^{\rm nat}$Ni and $^{\rm nat}$Mo targets with a diameter of 8 mm and mass of ~80 and 60 mg, respectively, were used in the experiment.
The isotopic composition of nickel is a mixture of five stable isotopes with an isotopic abundance: 
$^{58}$Ni -- 0.68077, $^{60}$Ni -- 0.26223, $^{61}$Ni -- 0.0114, $^{62}$Ni -- 0.03634, $^{64}$Ni -- 0.00926. For $^{100}$Mo in our calculations, we have used the percentage value of isotope abundance equal to 9.63\% (as in \cite{48}). The admixture of other elements in the targets did not exceed 0.1\% by weight. 

The $^{\rm nat}$Ni target and the $^{\rm nat}$Mo target-monitor were exposed to simultaneous radiation to bremsstrahlung radiation. The range of bremsstrahlung end-point energies $E_{\rm{\gamma max}}$ was from 35 to 94 MeV; the energy step was $\sim$~5~MeV.
The time of irradiation, $t_{\rm irr}$, was 30 min for each energy $E_{\rm{\gamma max}}$. The measurement time, $t_{\rm meas}$, of the residual $\gamma$-activity spectrum from the $^{\rm nat}$Mo targets was 30 min and in the case of  $^{\rm nat}$Ni targets was 20--95~hours.
The nickel targets were pre-cooled for $t_{\rm cool}$ = 32--53 days. Additionally, the induced activity spectra were measured for the $^{\rm nat}$Ni  targets irradiated with   $E_{\rm{\gamma max}}$ = 85.35, 89.25, and 94.0 MeV after a cooling period of 194--224 days.
The program InterSpec V.1.0.12 \cite{InterSpec} was used to process the spectra and obtained the full absorption peak areas $\triangle A$ for the relevant $\gamma$-ray transitions.

The bremsstrahlung spectra of electrons were calculated by using the GEANT4.9.2 toolkit  \cite{48,49,50}. Additionally, the bremsstrahlung flux was monitored against the
$^{100}$Mo($\gamma$,n)$^{99}$Mo reaction yield. This procedure was performed by comparing the experimentally obtained flux-averaged cross-section $\langle{\sigma(E_{\rm{\gamma max}})}\rangle$  values with the computation data. The monitoring procedure has been described in a great detail  in  \cite{a11,Ku,44}.

The parameter values of the  
${^{\rm nat}\rm{Ni}}(\gamma,{\rm p} x\rm n)^{58 \rm g,m}\rm{Co}$ reactions
according to the spectroscopic data from \cite{Live Chart} are listed in Table~\ref{table}. In addition, data are given for the \\ ${^{\rm nat}\rm{Ni}}(\gamma,x\rm n)^{56}$Ni reaction
which competes with \\ ${^{\rm nat}\rm{Ni}}(\gamma,{\rm p} x\rm n)^{58 \rm g}\rm{Co}$, and data for the monitor reaction \\ $^{100}$Mo($\gamma$,n)$^{99}$Mo.

\begin{table*}[]
	\begin{center}
		\caption{ Nuclear spectroscopic data of the nuclei-products from photonuclear reactions on $^{\rm nat}$Ni and $^{100}$Mo adopted from \cite{Live Chart}: spin $J$, parity $\pi$, half-life $T_{1/2}$ of the studied
			nucleus; $E_{\gamma}$ and $I_{\gamma}$ are the energies and intensities of the relevant $\gamma$-ray transitions; $E_{\rm{thr}}$ denotes the threshold energy of the contributing reaction (according to TALYS1.96 \cite{8,9}).}
		\label{table}  
		\begin{tabular}{ccccccc}
			\hline \hline\noalign{\smallskip}
			Nucleus ($J^{\pi}$) & $T_{1/2}$ & Decay mode (\%) & $E_{\gamma}$,~keV & $I_{\gamma}$, \% & 	\begin{tabular}{c} Contributing  \\ reactions 
			\end{tabular}  &$E_{\rm{thr}}$,~MeV
			\\ \noalign{\smallskip}\hline\noalign{\smallskip}
			$^{58 \rm m}$Co (5$^+$)  & 9.10(9) h & IT (100) & 24.889(21) & 0.0397(6) & 
			\begin{tabular}{c} $^{60}$Ni($\gamma$,pn)  \\ $^{61}$Ni($\gamma$,p2n) \\ $^{62}$Ni($\gamma$,p3n) \\ $^{64}$Ni($\gamma$,p5n)
			\end{tabular} &
			\begin{tabular}{c} 20.02  \\ 27.84 \\ 38.44 \\ 54.95
			\end{tabular}
			\\ \noalign{\smallskip}\hline\noalign{\smallskip}		
			
			$^{58 \rm g}$Co (2$^+$)  & 70.86(6) d & EC/$\beta^+$ (100) & 810.7593(20) & 99.45(1) & 
			\begin{tabular}{c} $^{60}$Ni($\gamma$,pn)  \\ $^{61}$Ni($\gamma$,p2n) \\ $^{62}$Ni($\gamma$,p3n) \\ $^{64}$Ni($\gamma$,p5n)
			\end{tabular} &
			\begin{tabular}{c} 19.99  \\ 27.81 \\ 38.41 \\ 54.92
			\end{tabular}
			
			\\ \noalign{\smallskip}\hline\noalign{\smallskip}
			$^{56}$Ni (0$^+$)  & 6.075(10) d & EC/$\beta^+$  (100) & \begin{tabular}{c} 158.38(3) \\ 811.85(3) \end{tabular} 
			& \begin{tabular}{c} 98.8(1.0) \\ 86.0(9) \end{tabular} & 
			\begin{tabular}{c} $^{58}$Ni($\gamma$,2n)  \\ $^{60}$Ni($\gamma$,4n) \\ $^{61}$Ni($\gamma$,5n) \\ $^{62}$Ni($\gamma$,6n)
			\end{tabular} &
			\begin{tabular}{c} 22.47  \\ 42.87 \\ 50.70 \\ 61.30
			\end{tabular}
			\\ \noalign{\smallskip}\hline\noalign{\smallskip}
			$^{99}$Mo (1/2$^+$)   & 65.924(6) h & $\beta^-$  (100) & 739.500(17) & 12.2(2)  &  $^{100}$Mo($\gamma$,n) & 8.29
			\\ 		\noalign{\smallskip}\hline\hline
		\end{tabular}	
	\end{center}
\end{table*}


\section{Calculation of the flux-averaged cross-sections and normalized reaction yield}
\label{sec:3} 

In this Section, details of the calculation of the cross-sections averaged by the bremsstrahlung flux are described. It will be further used to compare the experimental results with the theoretical estimates. 

The following equation can be used to calculate the experimental flux-averaged cross-section:
\begin{equation} 
	\langle{\sigma(E_{\rm{\gamma max}})}\rangle = 
	\frac{\lambda \  \triangle A \   {\rm{\Phi}}^{-1}(E_{\rm{\gamma max}})}{N_n I_{\gamma} \ \varepsilon \ (1-e^{-\lambda t_{\rm{irr}}})\ e^{-\lambda t_{\rm{cool}}}\ (1-e^{-\lambda t_{\rm{meas}}})},
	\label{form1}
\end{equation}

where $\triangle A$ denotes the number of counts in the full absorption peak corresponding to the $\gamma$-ray transition associated with the decay of the studied nucleus, $\lambda$ is the decay constant \mbox{($\rm{ln}2/\textit{T}_{1/2}$)}, $N_{\rm n}$ is the number of target nuclei, $I_{\gamma}$ is the absolute intensity of the relevant $\gamma$-ray transition, $\varepsilon$  is the absolute detection efficiency, $t_{\rm irr}$, $t_{\rm cool}$ and $t_{\rm meas}$ are the irradiation time, cooling time and measurement time, respectively. The quantity defined as ${\rm{\Phi}}(E_{\rm{\gamma max}}) = {\int\limits_{E_{\rm{thr}}}^{E_{\rm{\gamma max}}}W(E,E_{\rm{\gamma max}})dE}$ is the integrated bremsstrahlung $\gamma$-flux in the energy range from the reaction threshold $E_{\rm{thr}}$ up to $E_{\rm{\gamma max}}$. 

The theoretical $\langle{\sigma(E_{\rm{\gamma max}})}\rangle$ values were estimated using the cross-sections $\sigma(E)$ from the TALYS1.96 code and the bremsstrahlung $\gamma$-flux $W(E,E_{\rm{\gamma max}})$ calculated in\\  GEANT4.9.2 by the equation: 
\begin{equation}\label{form2}
	\langle{\sigma(E_{\rm{\gamma max}})}\rangle = 
	\frac{\int\limits_{E_{\rm{thr}}}^{E_{\rm{\gamma max}}}\sigma(E) \ W(E,E_{\rm{\gamma max}}) \ dE}{\int\limits_{E_{\rm{thr}}}^{E_{\rm{\gamma max}}}  W(E,E_{\rm{\gamma max}}) \ dE}.
\end{equation}

To calculate the cross-section of the  $^{\rm nat}$Ni($\gamma$,p$x$n)$^{58}$Co reaction, it is necessary to take into account that the production of the $^{58}$Co nucleus will occur on 4 stable nickel isotopes ($^{60}$Ni, $^{61}$Ni, $^{62}$Ni, and $^{64}$Ni). Therefore, Eq.~\ref{form2} is transformed into a sum of the flux-averaged cross-sections $\langle{\sigma(E_{\rm{\gamma max}})}\rangle_i$  for the production of the nucleus on each stable isotope $i$, which are calculated with their respective reaction threshold energy $E^i_{\rm thr}$:
\begin{equation*}\label{form1b}
	\langle{\sigma(E_{\rm{\gamma max}})}\rangle = 
	\sum^m_{i=1}	\frac{  \int\limits_{E^i_{\rm{thr}}}^{E_{\rm{\gamma max}}} A_i \ \sigma_i(E) \ W(E,E_{\rm{\gamma max}}) \ dE}{{\int\limits_{E^{i}_{\rm{thr}}}^{E_{\rm{\gamma max}}} W(E,E_{\rm{\gamma max}}) \ dE}},       (2\rm a)
\end{equation*}
where $\sigma_i(E)$ is the cross-section for the production of the $^{58}$Co nucleus on the $i$-th isotope with isotopic abundance $A_i$, $m$ is the number of stable isotopes of a natural element, on which an investigated nuclide can be formed.

Due to non-existence of the experimental partial reaction cross-sections for production of  $^{58}$Co on the four stable Ni isotopes, therefore we cant use results of calculation by Eq.~2a for compare with experimental data. 
We need to rewrite Eq.~2a so that for each $i$-th reaction will use the same integral flux, the bremstrahlung flux integrated from minimal reaction threshold:
\begin{equation*}\label{form1b}
	\langle{\sigma(E_{\rm{\gamma max}})}\rangle = 
	\frac{ \sum^m_{i=1} \int\limits_{E^i_{\rm{thr}}}^{E_{\rm{\gamma max}}} A_i \ \sigma_i(E)  \ W(E,E_{\rm{\gamma max}}) \ dE}{{\int\limits_{E^{\rm min}_{\rm{thr}}}^{E_{\rm{\gamma max}}} W(E,E_{\rm{\gamma max}}) \ dE}},       (2\rm b)
\end{equation*}
where $E^{\rm min}_{\rm thr}$ is the reaction threshold with the minimum energy. In the case of photoproduction of $^{58}$Co on $^{\rm nat}$Ni, the value of $E^{\rm min}_{\rm thr}$ = 19.99~MeV. 

In Eq.~2a the real integral bremsstrahlung fluxes for each $i$-th reaction with the corresponding threshold energy were taken into account. In the Eq.~2b flux-averaged cross-section was calculated using the "effective" integrated flux. Therefore, this leads to a difference in the results of calculating the flux-averaged cross-sections $\langle{\sigma(E_{\rm{\gamma max}})}\rangle$  according to Eq.~2a and 2b. This difference depends on the percentage of isotopes $A_i$, the threshold energy values of the $i$-isotope reactions $E^i_{\rm thr}$ and the minimum reaction threshold  $E^{\rm min}_{\rm thr}$, and also on the parameter values of the theoretical model  (for example, options $LD$ and $GSF$). 

To calculate the contribution of the reaction on the $k$-th isotope to the total production of the $^{58}$Co nucleus on natural nickel, the normalized yield formula is used, which can be written as follows:
\begin{equation}\label{form3}
	Y_k(E_{\rm{\gamma max}}) = \frac
	{\int\limits_{E^k_{\rm{thr}}}^{E_{\rm{\gamma max}}} A_k \ \sigma_k(E) \  W(E,E_{\rm{\gamma max}}) \ dE}
	{\sum^4_{i=1} \int\limits_{E^i_{\rm{thr}}}^{E_{\rm{\gamma max}}} A_i \ \sigma_i(E) \  W(E,E_{\rm{\gamma max}}) \ dE},
\end{equation}
where $\sigma_k(E)$ is the cross-section for the production of the $^{58}$Co nucleus on the $k$-th isotope with isotopic abundance $A_k$. 

\section{Results and discussion}
\label{sec:4}
\subsection{Calculated $\sigma(E)$ and $\langle{\sigma(E_{\rm{\gamma max}})}\rangle$  cross-sections}
\label{sec:4a}


In this work, the TALYS code \cite{8,9} with different level density models ($LD$) and gamma strength functions ($GSF$) were used to calculate the theoretical cross-sections $\sigma(E)$ for the studied reactions.

There are three phenomenological level density models and three options for microscopic level densities:

$LD 1$: Constant temperature + Fermi gas model, introduced by Gilbert and Cameron. 

$LD 2$: Back-shifted Fermi gas model.  

$LD 3$: Generalized superfluid model.

$LD 4$: Microscopic level densities (Skyrme force) from Goriely’s tables.

$LD 5$: Microscopic level densities (Skyrme force) from Hilaire’s combinatorial tables. 

$LD 6$: Microscopic level densities based on temperature-dependent Hartree-
Fock-Bogoliubov calculations using the Gogny force from Hilaire’s combinatorial tables.

The shape of the excitation function curve mainly depends on the gamma strength functions. There are nine gamma strength functions available in the TALYS1.96 code, namely:

$GSF 1$: Generalized Lorentzian of Kopecky and Uhl.

$GSF 2$: Generalized Lorentzian of Brink and Axel.

$GSF 3$: Skyrme-Hartree-Fock + BCS approximation.

$GSF 4$: Skyrme-Hartree-Fock-Bogoliubov model with quasiparticle random-phase approximation (QRPA).

$GSF 5$: Hybrid model of Goriely.

$GSF 6$: Temperature-dependent Skyrme-Hartree-Fock-Bogoliubov model with QRPA.

$GSF 7$: Temperature-dependent Relativistic Mean Field model.

$GSF 8$: The Gogny-Hartree-Fock-Bogoliubov model with QRPA, based on the D1M
version of the Gogny force.

$GSF 9$: Simplified Modified Lorentzian.

\begin{figure*}[]
	\resizebox{1.0\textwidth}{!}{%
		\includegraphics{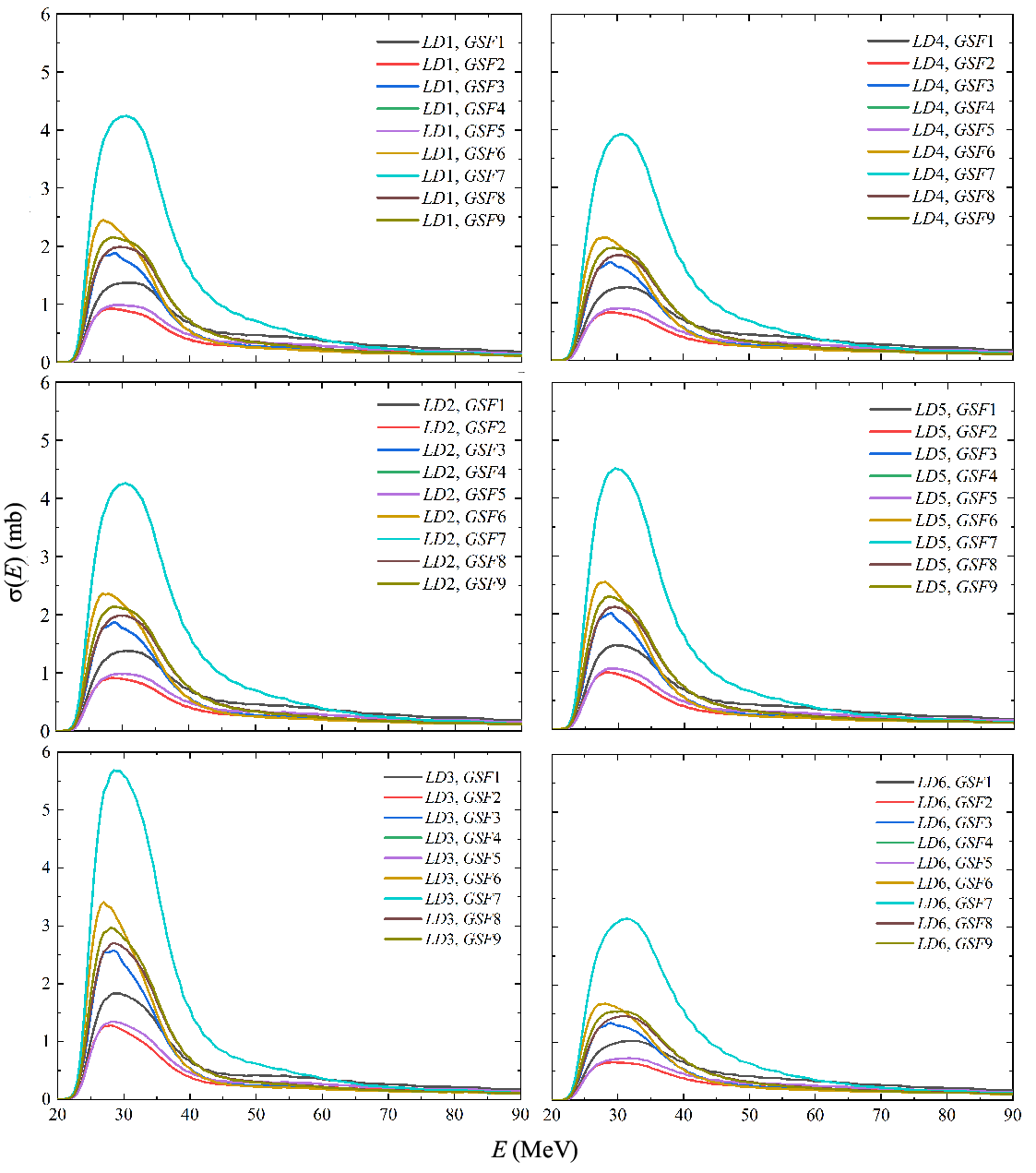}}
\caption{Cross-sections $\sigma(E)$ for the $^{\rm nat}$Ni($\gamma$,p$x$n)$^{58}$Co reaction calculated with the  TALYS1.96 code for six level density models $LD$ and nine gamma strength functions $GSF$.}
\label{fig2}
\end{figure*}

\begin{figure}[t]
\resizebox{0.49\textwidth}{!}{%
	\includegraphics{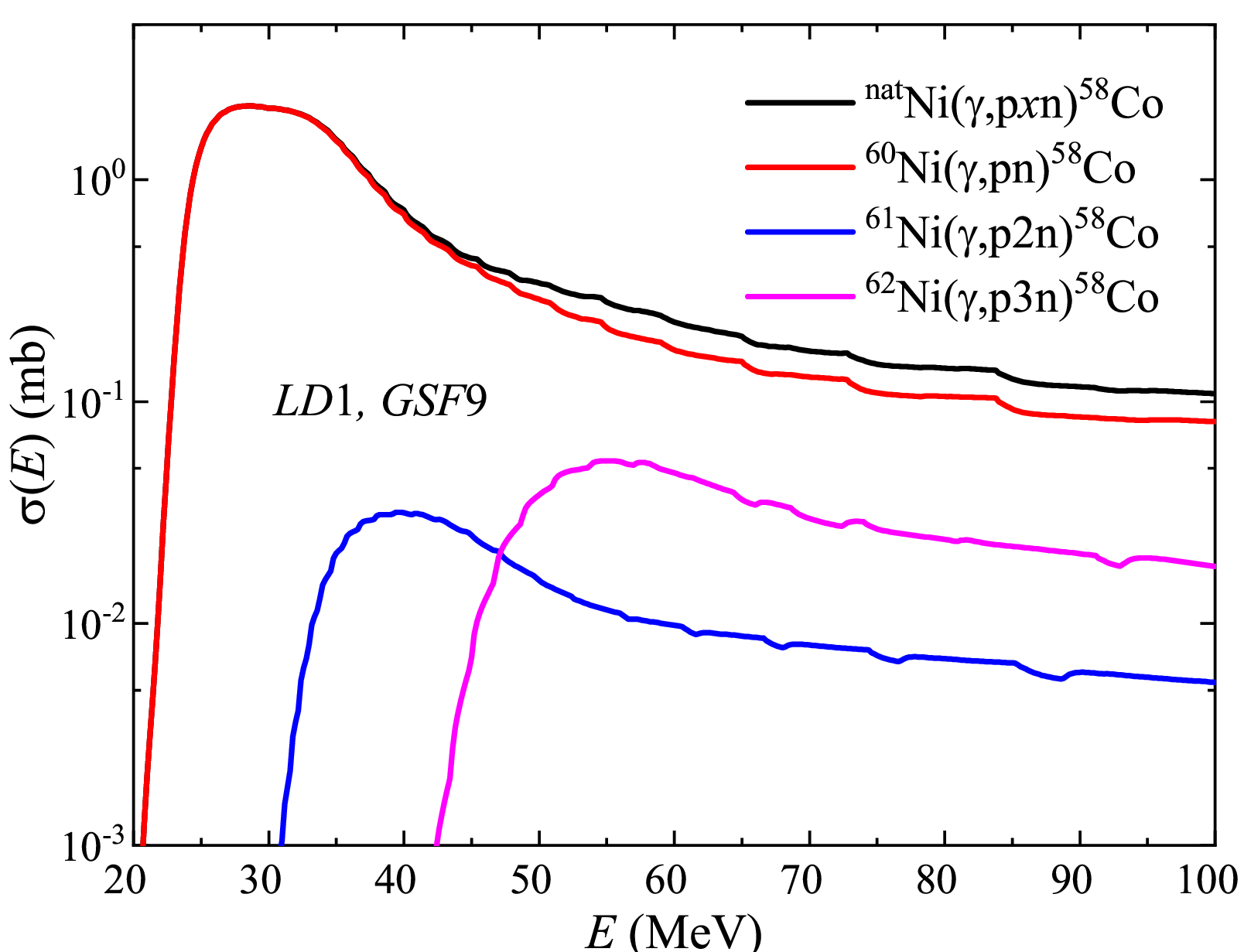}}
\caption{Cross-sections of photoproduction of the $^{58}$Co nucleus on nickel isotopes (taking into account the isotopic abundance) and on $^{\rm nat}$Ni, calculated in the TALYS1.96 code for the default parameters ($LD1$, $GSF9$).}
\label{fig3}
\end{figure}

Changing the parameters $LD$ 1--6 and $GSF$ 1--9 leads to 54 different results of the theoretical value of $\sigma(E)$, and the observed spread of cross-sections reaches $\sim$~8 times (see Fig.~\ref{fig3}). The minimum value was obtained for the case of a combination of parameters $LD$6 and $GSF$2, and the maximum -- $LD$3 and $GSF$7. Note that the combination of parameters $LD$1, $GSF$9 are used by default for calculation in TALYS1.96.

Figure~\ref{fig2} shows the calculated cross-sections $\sigma(E)$ of the production of the $^{58}$Co nucleus in photonuclear reactions on stable nickel isotopes, taking into account the isotopic abundances. 
The cross-section value of the production of the nuclide on $^{\rm nat}$Ni is given as a sum of cross-sections of reactions on 4 stable Ni isotopes.
As can be seen in Fig.~\ref{fig2}, the $^{60}$Ni($\gamma$,pn) reaction is the dominant reaction.  To  determine the contribution of the $\gamma$ + $^{60}$Ni reaction to the total production of $^{58}$Co, the normalized yield $Y_{60}$ was calculated according to Eq. 3. Using the default set of parameters in the TALYS1.96 code, this contribution was determined as 99.99\% at $E_{\gamma\rm max}$ = 35.55~MeV, while decreasing to 96.9\% at $E_{\gamma\rm max}$ = 94~MeV. Therefore, the accuracy of calculating the  cross-section of the $^{\rm nat}$Ni($\gamma$,p$x$n)$^{58}$Co reaction  is completely determined by the accuracy of calculating the  $^{60}$Ni($\gamma$,pn) reaction cross-section.

In Fig.~4, theoretical estimates of flux-averaged cross-section $\langle{\sigma(E_{\rm{\gamma max}})}\rangle$ for different combinations of $LD$ and $GSF$ parameters are shown.
These calculations were performed according to Eq. 2b and $E^{\rm min}_{\rm thr}$ = 19.99~MeV. The "effective" integrated flux from $E^{\rm min}_{\rm thr}$ is greater than the real flux, leading to negligible underestimation of the  \\ $\langle{\sigma(E_{\rm{\gamma max}})}\rangle$ values. This difference can be estimated using theoretical estimates of the reaction cross-sections and Eq.~2a to Eq.~2b ratio products. 
The magnitude of the difference smoothly increases with increasing bremsstrahlung end-point energy from 0\% up to  2.4\% in the investigated energy region. 

Note, in the studied energy range the value of $Y_{60}(E_{\gamma\rm max})$ and the Eq.~2a/Eq.~2b ratio is only slightly depended on the change of model parameters.

\begin{figure}[]
\resizebox{0.49\textwidth}{!}{%
	\includegraphics{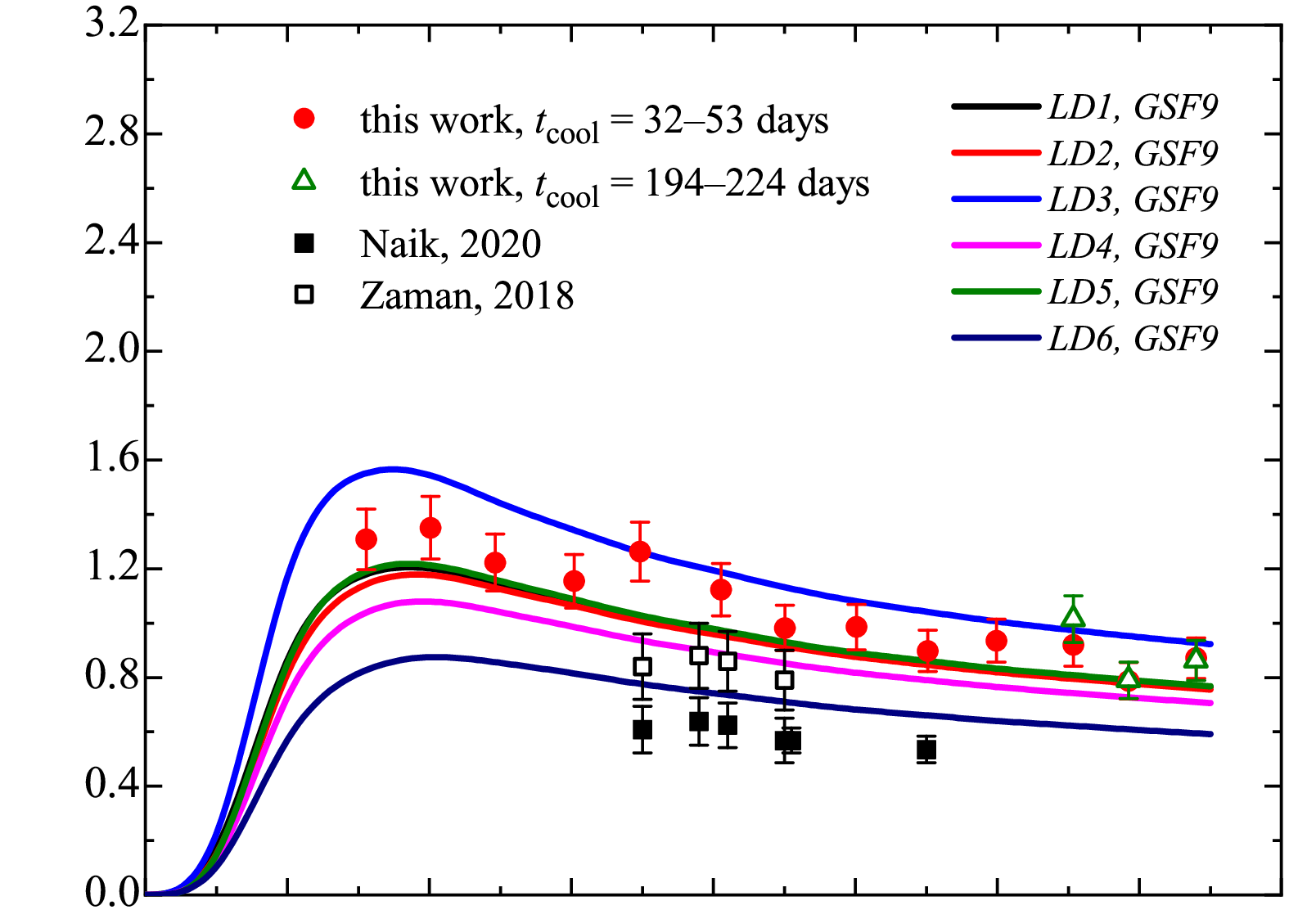}}
	\resizebox{0.49\textwidth}{!}{%
		\includegraphics{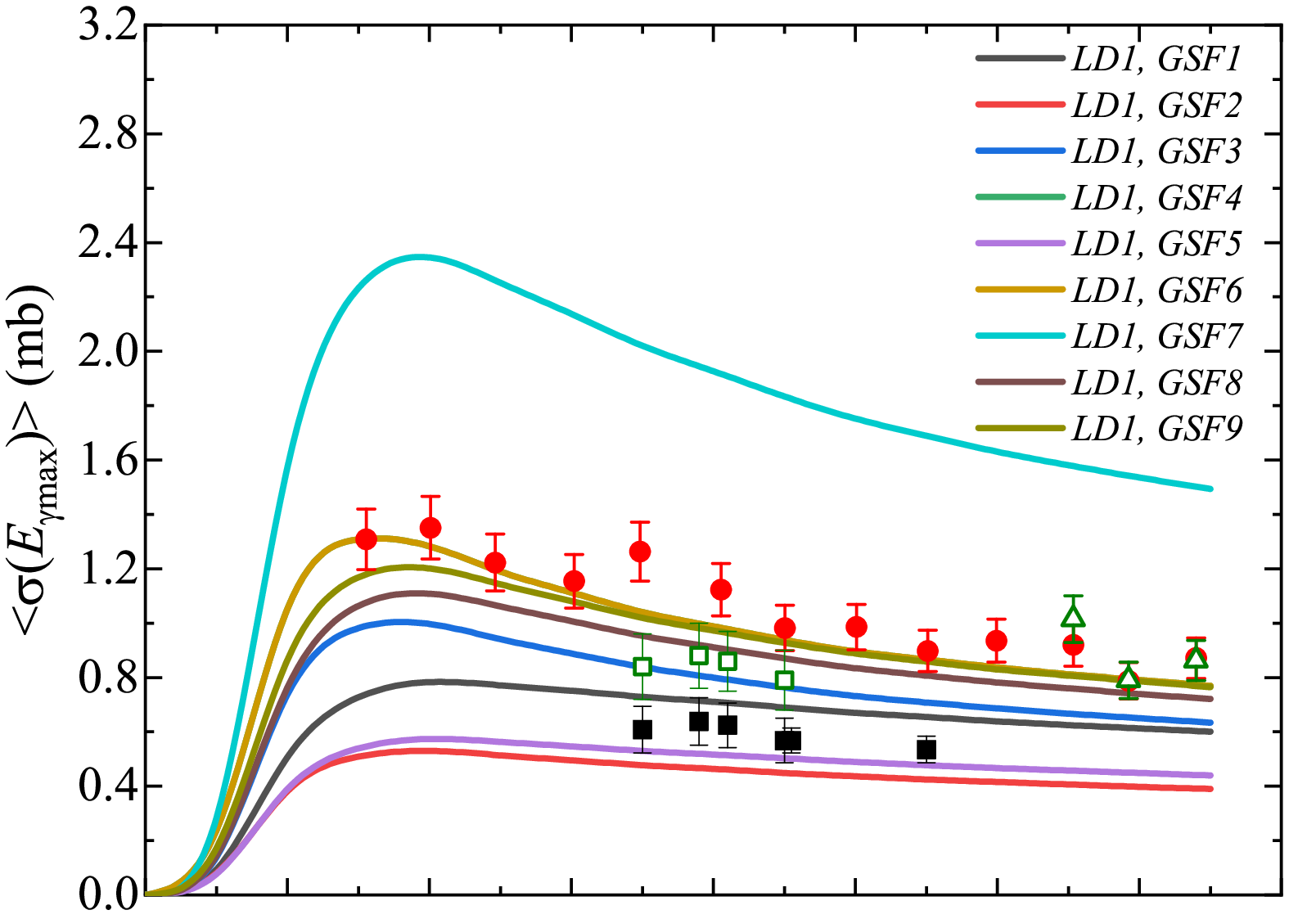}}
	\resizebox{0.49\textwidth}{!}{%
	\includegraphics{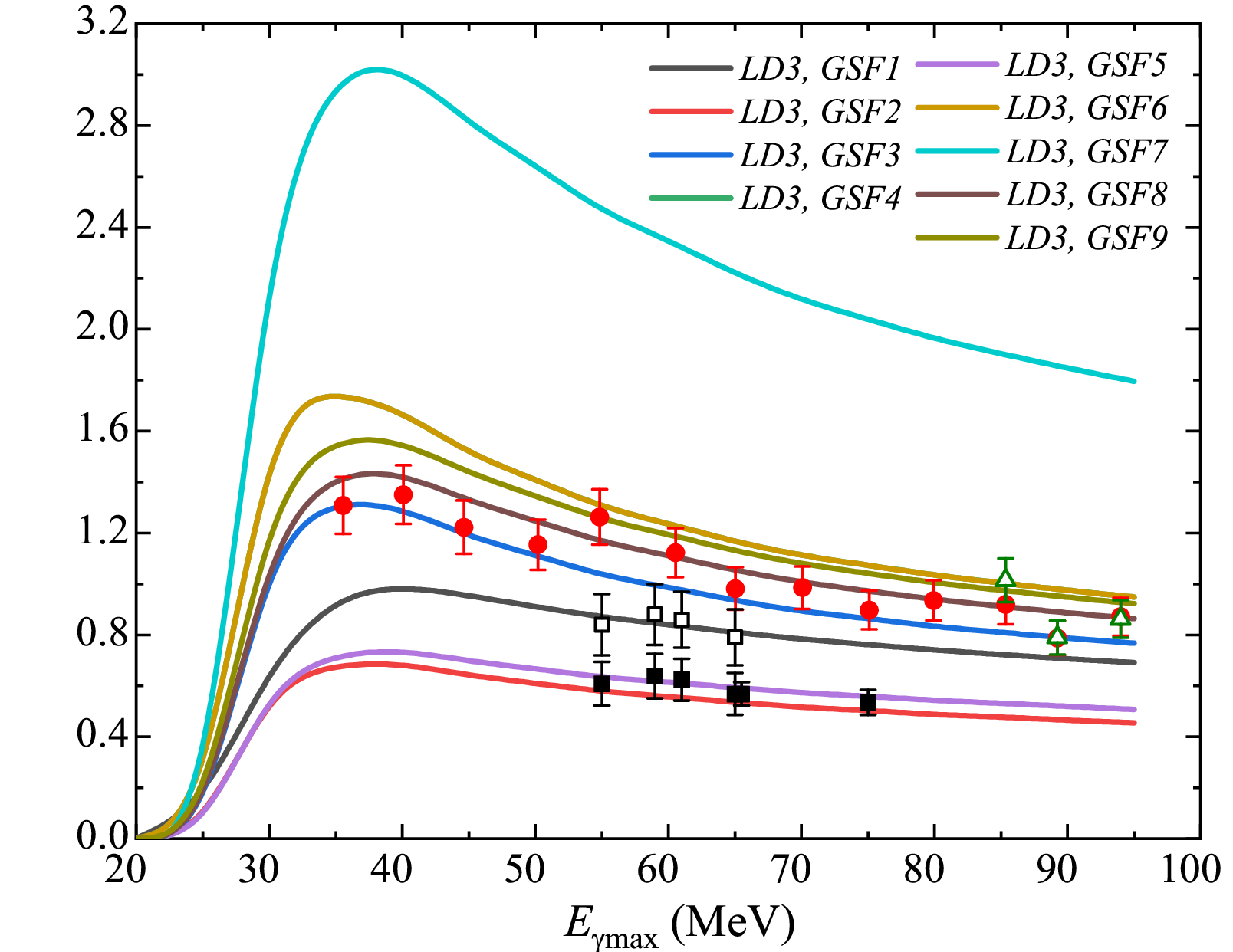}}
	\caption{The bremsstrahlung flux-averaged cross-sections $\langle{\sigma(E_{\rm{\gamma max}})}\rangle$ for the ${^{\rm nat}\rm{Ni}}(\gamma,{\rm p} x\rm n)^{58}$Co reaction. The experimental data: red circles and green triangles -- our data for $t_{\rm cool}$ =  32--53 days and 194--224 days, respectively; black empty and full circles -- data from Zaman et al. \cite{34} and Naik et al. \cite{35}, respectively. The theoretical estimates are shown as solid curves, where each colour represents different $LD$ and $GSF$ combinations used in the calculation. }
	\label{fig4}
\end{figure}

\subsection{Experimental values of bremsstrahlung flux-averaged cross-section $\langle{\sigma(E_{\rm{\gamma max}})}\rangle$}
\label{sec:4b}

As a result of irradiation with high-energy $\gamma$-quanta, multiparticle nuclear  
($\gamma$,$y$p$x$n) reactions occur in the $^{\rm nat}$Ni targets.  As a direct consequence, this leads to a high complexity of the measured  $\gamma$-ray spectra. Figures 5a and 5b show the spectra of $\gamma$ rays measured after irradiation of one $^{\rm nat}$Ni target, where target cooling times $t_{\rm cool}$ = 2925610 s and 19353600 s, respectively, were used.

\begin{figure*}[]
	\resizebox{1.0\textwidth}{!}{%
		\includegraphics{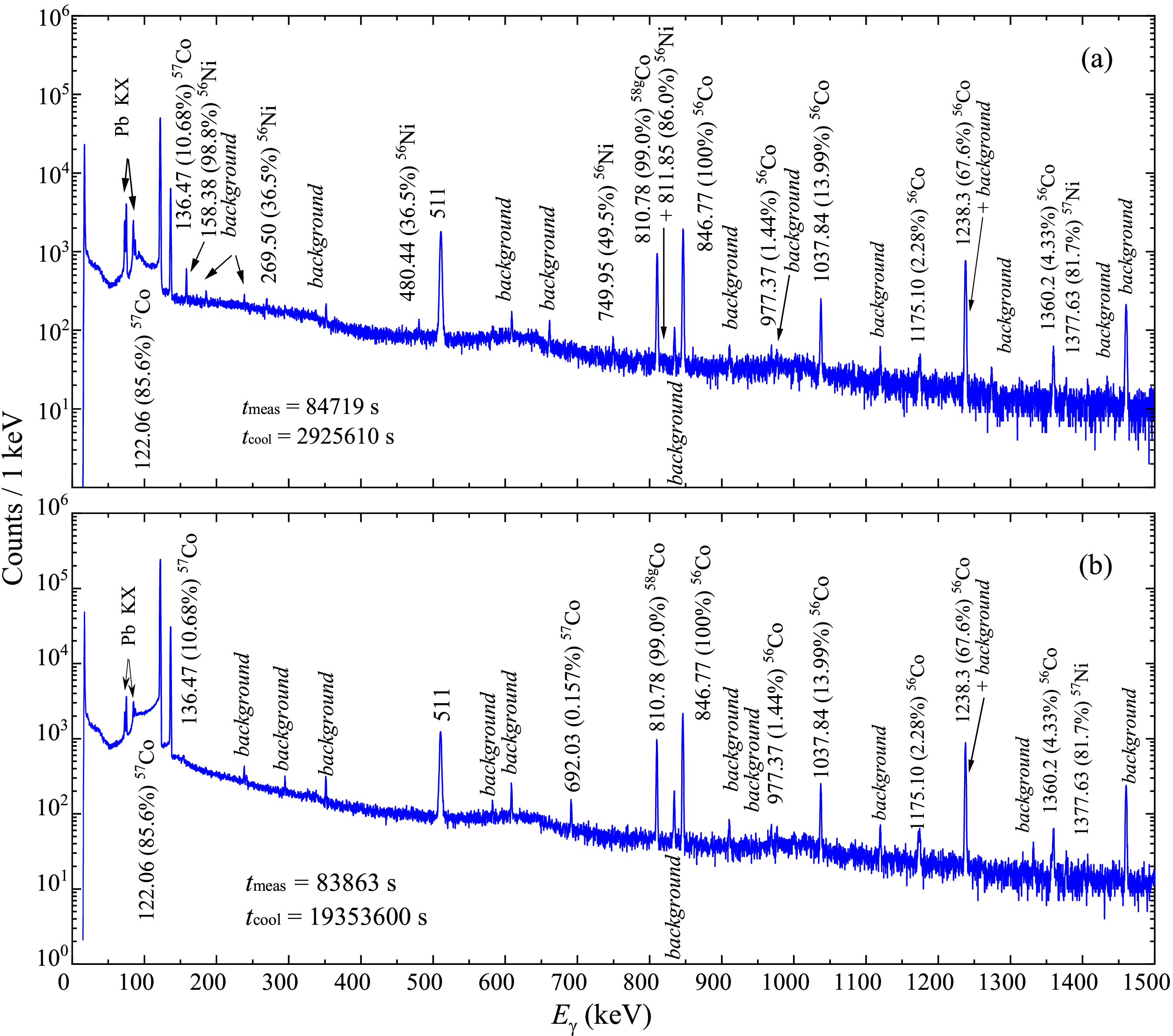}}
	\caption{Energy spectra of $\gamma$ rays measured by the HPGe detector, and emitted by the $^{\rm nat}$Ni target after the bremsstrahlung flux exposure time of 30 min with $E_{\rm{\gamma max}}$ = 89.25~MeV. Relevant peaks are labeled with their energies and intensities (shown in parentheses). 
		In panels (a), the cooling and measuring times were $t_{\rm cool}$ = 2925610 s, $t_{\rm meas}$ = 84719 s, respectively, while data shown in panel (b) were collected with time conditions $t_{\rm cool}$ = 19353600 s and $t_{\rm meas}$ = 83863 s.  }
	\label{fig5}
\end{figure*}

The $^{58}$Co nucleus has a known isomeric state at an excitation energy of 24.89 keV \cite{Live Chart}. In the reaction, both the $^{58}$Co ground and isomeric states  are populated. Due to the nature of the isomeric state and the transition depopulating it, it was not possible to determine the cross-section for the production of $^{58 \rm m}$Co in the experiment. Since the isomeric state feeds the ground state via a low-energy transition with a half-life of 9.10(9) h \cite{Live Chart}, and the total cooling time of the irradiated target was minimum 32 days, thus allowing the  $^{58 \rm m}$Co nuclei deexcite to their gound state, the total cross-section for the production of  $^{58}$Co was measured instead.

To determine the experimental values of $\langle{\sigma(E_{\rm{\gamma max}})}\rangle$ for the $^{\rm nat}$Ni($\gamma$,p$x$n)$^{58}$Co reaction, information about the 810.76 keV transition with an absolute intensity of 99.45\% (see Table I) was used. However, transition with a similar energy, 811.85 keV and an absolute intensity of 86.0\% is as well present in the data as a result of the decay of $^{56}$Ni produced in the $^{\rm nat}$Ni($\gamma$,$x$n) reactions. This causes the peak at ~811 keV to be a doublet, artificially increasing the intensity of the 810.76 keV transition.
To calculate the magnitude of this contribution, the induced activity of the $^{56}$Ni nucleus was measured for the 158.38~keV transition. Then, this value was recalculated to the expected activity for the 811.85~keV transition, taking into account the intensities $I_{\gamma}$(158.38) = 98.8\% and $I_{\gamma}$(811.85) = 86.0\%, and the detection efficiency $\varepsilon$. The extracted values \\
$\triangle A$(811.85) = $\triangle A(158.38) \times $\\ ($I_{\gamma}(811.85) \times \varepsilon$(811.85))/($I_{\gamma}(158.38) \times \varepsilon$(158.38)) \\
were subtracted from the observed amounts \\ $\triangle A$(810.78+811.85) in all measured spectra with $t_{\rm cool}$ = 32--53 days. The magnitude of this contribution depends on the cooling time and was determined to be 0.9--8.5\%.

To check what effect the presence of the 811.85 keV transition has on the resulting flux-averaged cross-section of the studied reaction, additional measurements were performed after a cooling time of $t_{\rm cool}$ = 194--224 days. With such a long cooling time, the peaks at energies of 158.38, 269.50, 480.44, and 749.95~keV, which correspond to the 
decay of the $^{56}$Ni nucleus, were absent from the induced activity spectra. Therefore, a 
contribution from $^{56}$Ni in the peak at $E_{\gamma}$ = 810.76~keV should also be 
absent (see, for example, Fig. 5b). As can be seen in Fig.~4, the data from the two types measurements agree with each other within the uncertainty limits.

The calculation of the experimental $\langle{\sigma(E_{\rm{\gamma max}})}\rangle$ was performed according to Eq.~1 using the bremsstrahlung flux from the reaction threshold with the minimum threshold energy $E^{\rm min}_{\rm thr}$ = 19.99~MeV. The experimental results 
for the  $^{\rm nat}$Ni($\gamma$,p$x$n)$^{58}$Co reaction determined in this work are summarized in Table~II and graphically shown in Fig.~4.

The accuracy of the experimentally determined cross-sections $\langle{\sigma(E_{\rm{\gamma max}})}\rangle$ in measurements was determined as square-root of quadratic sum of statistical and systematical errors. The statistical error in the observed $\gamma$-activity is mainly due to statistics in the full absorption peak of the corresponding $\gamma$-transition, which varies within 1.0--1.5\%.

The systematic errors are due to the following uncertainties: 1) exposure time and the electron current $\approx$~0.5\%; 2) detection efficiency of the HPGe detector $\approx$~2--3\%, which is generally associated with the $\gamma$ radiation 
source uncertainty and the choice of the fitting function; 3) half-life of the reaction products and the absolute intensity $I_{\gamma}$ of the analyzed $\gamma$-ray transitions; 4) the number of target atoms $\approx$~0.2\%; 5) normalization of the 
experimental data to the yield of the monitoring reaction $^{100}$Mo($\gamma$,n)$^{99}$Mo made up to 5 \%; 6) contribution of the 811.85 keV transition from the $^{56}$Ni nucleus to the full absorption peak at $\sim$~811.0~keV was found to be in the range of 0.9 to 8.5\%. This resulted in an additional error of less than 1\%.

Two sets of experimental flux-averaged cross-sections $\langle{\sigma(E_{\rm{\gamma max}})}\rangle$ values for photonuclear $^{\rm nat}$Ni($\gamma$,p$x$n)$^{58}$Co reaction have been published so far, these include studies by Zaman et al. \cite{34} and Naik et al. \cite{35}. 
Note that $\langle{\sigma(E_{\rm{\gamma max}})}\rangle$ from these works differ by a correction, that take into account the contribution of the activity of the $^{56}$Ni nucleus for the $\gamma$-ray transition with an energy of $E_{\gamma}$ = 810.78~keV. As can be seen from Fig. 4, the magnitude of this 
contribution can have a significant impact on the cross-sections $\langle{\sigma(E_{\rm{\gamma max}})}\rangle$ for the $^{\rm nat}$Ni($\gamma$,p$x$n)$^{58}$Co reaction.

Comparison of the $\langle{\sigma(E_{\rm{\gamma max}})}\rangle$ values for the \\ $^{\rm nat}$Ni($\gamma$,p$x$n)$^{58}$Co reaction has shown that the cross-section values from this work are larger than those reported in \cite{34,35}. Remarkably, our results on reaction cross-sections from the study of $^{\rm nat}$Ni($\gamma$,p$x$n)$^{55,56,57}$Co reactions are in a good agreement with those reported in \cite{34,35}.

As can be seen in Figs. 3 and 4, theoretical calculations for different $LD$ and $GSF$ parameter combinations are scattered all over the plane, and in some cases they are different by a factor of 8.
There are a large number of theoretical estimates near the experimental data found in the work, which makes the procedure for selecting the optimal theoretical model practically impossible. Since the results of comparison by the least squares method have close values, it is impossible to choose one of the calculation options. 
The comparison showed that our measured flux-averaged cross-sections for studied reaction are in satisfactory agreement with several calculation estimates within the TALYS1.96 code ($LD$1 + $GSF$ 6,9; $LD$3 + $GSF$ 3,8,9).

\begin{table}[h]
	\caption{\label{tab1}Experimental total flux-averaged cross-section $\langle{\sigma(E_{\rm{\gamma max}})}\rangle$ [mb]	for the $^{\rm nat}$Ni($\gamma$,p$x$n)$^{58}$Co and $^{60}$Ni($\gamma$,pn)$^{58}$Co reactions.}
	\centering
	\begin{tabular}{ccc}
		\hline\hline\noalign{\smallskip}
		$E_{\rm{\gamma max}}$,~MeV & $\;\;\;\;$ $^{\rm nat}$Ni($\gamma$,p$x$n)$^{58}$Co &  $\;\;\;\;$ $^{60}$Ni($\gamma$,pn)$^{58}$Co* \\ \noalign{\smallskip}\hline\noalign{\smallskip}	
		35.55 & 1.31(11) & 4.98(43) \\ 
		40.10 & 1.35(12) & 5.13(44) \\ 
		44.65 & 1.22(10) & 4.63(39) \\ 
		50.20 & 1.15(10) & 4.35(37) \\ 
		54.85 & 1.26(11) & 4.75(41) \\ 
		60.55 & 1.12(10) & 4.20(36) \\ 
		65.05 & 0.98(8) & 3.66(31) \\ 
		70.10 & 0.99(8) & 3.66(31) \\ 
		75.10 & 0.90(8) & 3.33(28) \\ 
		79.95 & 0.94(8) & 3.47(29) \\ 
		85.35 & 0.92(8) & 3.41(29) \\ 
		89.25 & 0.79(7) & 2.91(25) \\ 
		94.00 & 0.87(7) & 3.22(27) \\ 
		\noalign{\smallskip}\hline\hline	  	
	\end{tabular} \\ 
	\footnotesize{* The flux-averaged cross-sections of the $^{60}$Ni($\gamma$,pn)$^{58}$Co reaction were estimated using the normalized reaction yield $Y_{60}(E_{\gamma \rm max})$ according Eq.~3  with combination of parameters $LD$1 and $GSF$9 and the $^{60}$Ni isotopic abundance (0.26223).}
\end{table}

\section{Conclusions}
\label{Concl}

In this work,  experimental values of total bremsstrahlung flux-averaged cross-sections $\langle{\sigma(E_{\rm{\gamma max}})}\rangle$ for the photonuclear reactions $^{\rm nat}$Ni($\gamma$,p$x$n)$^{58}$Co and $^{60}$Ni($\gamma$,pn)$^{58}$Co in the range of end-point energies $E_{\rm{\gamma max}}$ = 35--94~MeV were determined. The experiment was performed using the \\ bremsstrahlung beam delivered by the NSC KIPT electron linear accelerator LUE--40. In the measurements, $\gamma$ activation method and $\gamma$-ray spectrometric techniques were used. 
The obtained $\langle{\sigma(E_{\rm{\gamma max}})}\rangle$ cross-sections and the experimental  data from the literature \cite{34,35} were compared. It was shown that our data were higher than those reported in Refs. \cite{34,35}. 

The theoretical values of $\langle{\sigma(E_{\rm{\gamma max}})}\rangle_{\rm th}$ were calculated using the partial cross-section $\sigma(E)$ from the TALYS1.96 code for the different level density models, gamma strength functions, and bremsstrahlung $\gamma$-flux calculated by the GEANT4.9.2 toolkit. A large number of curves with close results makes it difficult to choose the optimal theoretical estimate for describing the experimental data. The comparison showed that the measured flux-averaged cross-sections for the $^{\rm nat}$Ni($\gamma$,p$x$n)$^{58}$Co reaction are in satisfactory agreement with several calculation estimates within the TALYS1.96 code.

\section*{Acknowlegment}
The authors would like to thank the staff of the linear electron accelerator LUE-40 NSC KIPT, Kharkiv, Ukraine, for their cooperation in the realization of the experiment.

This work was supported by the Slovak Research and Development Agency under No. APVV-20-0532, and the Slovak grant agency VEGA (Contract No. 2/0175/24). Funded by the EU NextGenerationEU through the Recovery and Resilience Plan for Slovakia under the project No. 09I03-03-V01-00069.

\section*{Declaration of competing interest}
The authors declare that they have no known competing financial interests or personal relationships that could have appeared to influence the work reported in this paper.

\end{document}